# WHERE TELESCOPES CANNOT (YET) SEE - THE MOON AS SEEN BY SCRIVEN BOLTON, ÉTIENNE TROUVELOT, LUCIEN RUDAUX, CHESLEY BONESTELL

**Angelo Adamo**

*INAF-Bologna Astronomical Observatory*

**ABSTRACT**

Scientific illustrations, thanks to the vision of great artists fascinated by astronomical research and astronautics, have provided us with an accurate depiction of the possible views which mankind will one day observed from locations other than our planet. In this talk I will pay homage to some of these geniuses who serve science, and underline the scientific, artistic, political, and social implications deriving from a wise use of space-art.

## 1. THE BEGINNING OF SPACE ART

In this paper I will discuss primarily the work by Chesley Bonestell (1888-1986), who studied as an architect, and was probably the most famous scientific illustrator of the 20th century. He specialised in creating astronomical images, and influenced through his work the collective imagination of entire generations, or even generated it *ex novo*.

But, before discussing his production, we should first place him in the context of his precursors (see Fig.1).

I would therefore start from the French illustrator and astronomer Etienne Trouvelot (1827-1895): he was active at a time when photography was already commonly used in several fields, including astronomy, but nevertheless he decided to use his undeniable artistic qualities to illustrate cosmic subjects with a richness of detail which, in his view, could only be caught by the human eye. The artist who immediately preceded Bonestell, and in some measure influenced his style, was the Briton Scriven Bolton (1883-1929). A Fellow of the *Royal Astronomical Society*, he was keen on both astronomy and painting, while not being a professional in either. The young American architect Bonestell, who had moved to Britain to work in a studio in London, also started writing for the *Illustrated London News* magazine, where he first saw Bolton's work. He later confessed to not really liking it. He is quoted as saying:

*He made annoying mistakes and could not paint mountains. It was partly because of his mistakes that I decided to indulge in space painting*".

After leaving London to return to America, Bonestell started working for the Hollywood film industry, painting *mattes*, artificial backgrounds used in film scenes to simulate landscapes and therefore avoid expensive transfers of entire casts to far locations. Working in this environment gave him not just financial security, but also the possibility to access sets outside working hours. He took this opportunity to study the effects produced by light from a distant source on spheres positioned so as to simulate planets and their satellites. To increase the scientific accuracy of his astronomical illustrations, and despite his not appreciating Bolton, he borrowed from him a smart technique which is still used by many *space artists*: he built plastic models of "alien" surfaces by copying terrestrial landscapes and views, to which, after photographing them, he added stars, craters, asteroids, … by painting them directly on the plates.

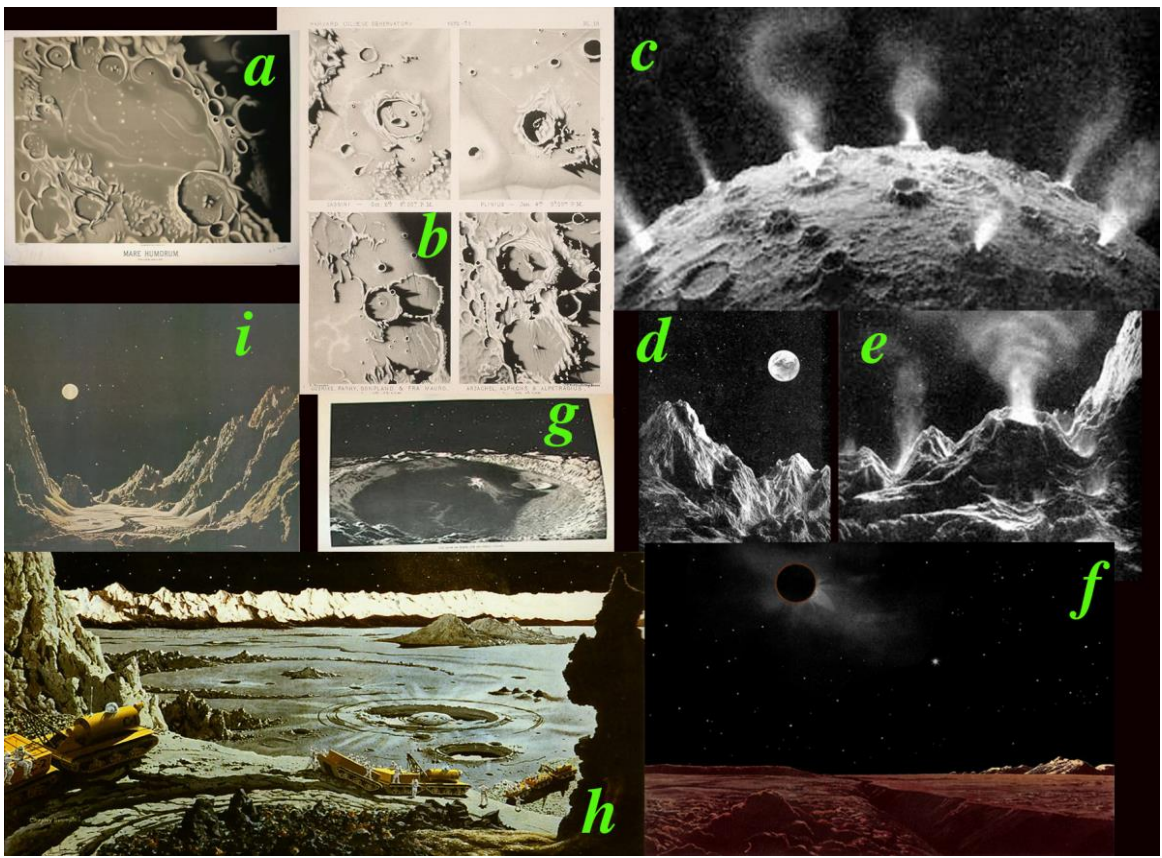

*Figure 1 – Comparison between different representations of the moon given by E. Trouvelot (a,b), S. Bolton (c, d, e), L. Rudaux (f, g) and C. Bonestell (h, i)*

## 2. BUILDING THE IMAGE OF THE MOON

This technique allowed Bonestell to become much appreciated artist in this subject, and attracted upon him the attention of NASA, just at the time when the Agency was preparing to send a man to the moon.

This established the official collaboration between NASA and Bonestell, who built the entire icongraphy which was used to divulge to a vast public the exciting start of the space era. This collaboration - with an institution which had the means to spread his work thoroughly, not just in the USA, but in the entire world (because of the importance of space exploration in those years) – influenced the public imagination to the extent that Bonestell's paintings became the standard idea of alien landscapes for everyone, whether they were scientists or not. The dramatic views of the moon, which he rendered as wide plains surrounded by high saw-tooth mountains; the loneliness of distant satellites hidden in the shadows of the giants Jupiter and Saturn; the threatening view of Mars in the background of its asteroidal satellite Phobos, … they were all so beautiful, evocative, and exciting, that they became the best allies of the propaganda produced by NASA and by the American government. This use of scientific illustrations can rightly be considered as one of the political, communicative, and technological strategies which reached their apex with the moon landing of 20 July 1969. Furthermore, a smart political rhetoric presented the event not as an exclusive American success, but as a symbolic victory of the Western world against the obscure enemy who was hiding beyond the Berlin wall, and who in 1957 had started launching "threatening" satellites into space to spy upon the rest of the world. But when the first *lunar orbiters* started observing our only natural satellite, they revealed that its surface was much more flat, dull, and sadly uninteresting than depicted by Bonestell, who had imagined a moon with an interesting geology formed by high, craggy mountains. If he had illustrated it as it really was, no taxpayer, probably, would have supported a mission to explore such a dull place. But one thing is certain: his was not a deliberate artistic deception. His convictions on the moon's look were probably rooted in a century dominated by romantic ideas, something which influenced him for the rest of his long life. Although he carefully observed the moon with his own telescope, he used the instrument as a projector, showing in the sky just what his fertile imagination wanted him to see. Just like his famous predecessor, Galileo Galilei (1564-1642), Bonestell was keen to impress his audience: consciously or not, the lack of a lunar flora and fauna led him to over-dramatize geological elements of the landscape such as mountains and rough surfaces, the only ones which could trigger the viewers' imagination through this emphasis. We recall what Battistini (2001) writes about Galilei's illustrations (Galilei, 1610):

*Perhaps Gombrich's conclusion in* Art and Illusion *also applies to him: the figurative experience is influenced by the mediation of pre-crystallized shapes, such as those inherited from Plutarch's selenic descriptions, or from the Copernican philosophical assumptions. But most certainly the illustrations in* Sidereus *no longer appear as a decorative frame conditioned by ornamental taste and pictorial complacency, but rather as a didactic guide raised to become an integral part of the scientific discourse. They are essential, because their geometric accuracy embodies the clarity of the written text, which is too much a subject, through the alphabet, of the mimics of the acoustic oral speech. One was surely much more impressed by the dark shadows emphasizing the mountains in his sketches – which left their sign on the painting style of the time - than by mere words.*

If we replace Galileo's name with Bonestell's or Bolton's, we understand why their illustrations conformed so well to the taste which characterized their childhood and the period in which they grew professionally.

But we could go even further and claim that, if Bonestell's illustrations achieved such a grip on the public, it was because, through those "pre-crystallizations" mentioned by Gombrich, the entire western world still felt the same romantic furor as Bonestell: it was this "sympathy", this harmony between his taste and that of the public, which led him to such a great success. To support this we can mention Lucien Rudaux (1874-1947), an expert astronomer and also a well-appreciated illustrator who specialized in astronomical subjects. His oeuvre reveals a great attention to what he could see through a telescope: the moon, Mars, and the other subjects he painted are nothing else but objective representations of how their surfaces actually appeared. The moon, in particular, a frequent subject of Rudeaux's, shows very wide, eroded plains, scarred by billions of years of impacts from asteroids, not filtered by an atmosphere, which as we know is absent from our satellite. Bonestell's view of Mars included the famous channels seen by Schiaparelli first and by Lowell later, those same channels which supported "scientifically" H.G Wells' Darwinian nightmares, later cleverly exploited by Orson Welles; while Rudaux's martian surface is just what it is: a bleak stony plain with not many narrative possibilities, under a thin sky which has just as little to suggest. Thus, despite being an erroneous representation, the images created by Bonestell can be credited for convincing the entire western world to converge towards that single ambitious, but also hugely expen-sive objective, which was the human exploration of the moon.

Rudaux nonetheless had some success, but neither NASA nor the public at large appeared to have really noticed his work, so that he never had a chance to create an aesthetic trend like that deriving from his American competitor's ilustrations. In hindsight, we must acknowledge the difficulty of seducing viewers with images which were so accurate in depicting a bleak, desert reality, devoid of mystery and intrigue. Instead, teenagers, children, adults and elderly people of all nationalities saw Bonestell's images of the moon and could immediately day-dream of being on our satellite, or wherever they wanted to be in the cosmos. They had the opportunity of observing from a different point of view, travelling in new, stimulating landscapes which could be explored and colonized. In hindsight, it becomes obvious that the appeal of Bonestell's large paintings came from his decision to paint terrestrial landscapes, which only in a second phase became exotic through the subtraction of elements such as trees, grass and air. This was followed by yet another phase, in which stars, craters, asteroids… were added. The viewers of Bonestell's paintings would find themselves in the cosmic elsewhere, while staying firmly rooted on our planet. Their subconscious would recognize some elements of earthly landscapes, which would keep them from feeling displaced, away from home, from family and from security; while the extraterrestrial elements would do the opposite, and make a viewer feel alienated and displaced from reality.

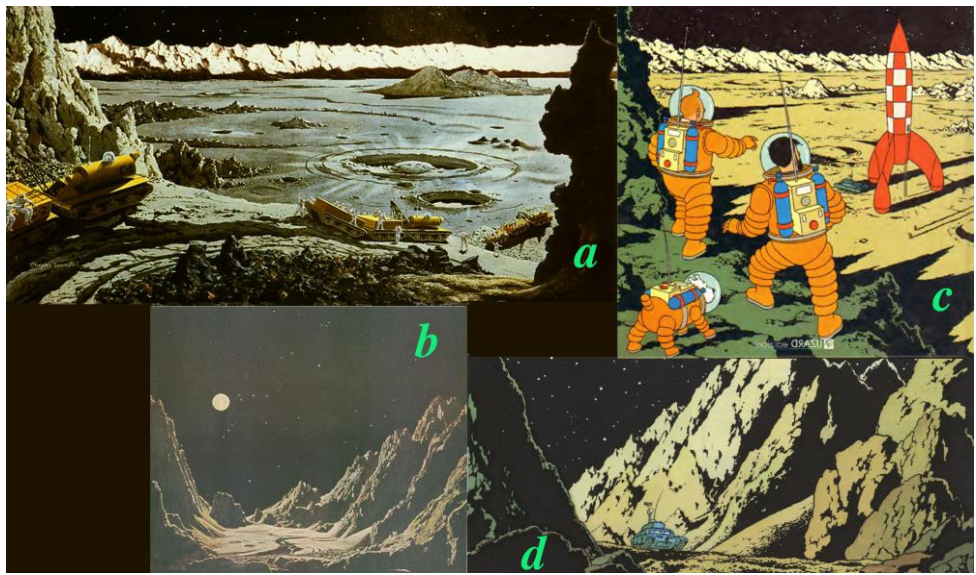

*Figure 2 - Comparison between the representation of the moon given by C. Bonestell (a, b) and Hergé (c, d)*

## 3. SPREAD OF BONESTELL'S VISION

In order to understand how meaningful the effect of this visionary's images must have been, and to comprehend their superiority to those of Rudaux (however excellent the latter), I submit to the reader's consideration what I found in my continuous wanderings among different astronomical illustrations. Just as a premise: as we know, in order to understand an entire historical period, we can execute a kind of cultural "core drilling" of various aspects of society, and thus comprehend what was communicated in that period and how it was spread. I chose to apply this analysis to the comic book world, because this is an excellent medium to understand the trends of western society. I found that some frames in the comic book "Destination Moon", written and illustrated by Hergé (1907-1983), the well-known creator of the even better-known character *Tin Tin*, were not just *inspired by*, but actually copied from paintings by Bonestell (see Fig. 2): a proof of the influence of his work on the western culture of early post-World War II years. And the influence of illustrators such as Boneswell on the film industry has already been mentioned earlier in this paper. They often were called to prepare the *storyboards* of films: these are preparatory drawings of key moments in the script, which allowed to plan the final visual result in the film. When many frames have to be drawn, the storyboard becomes *de facto* a comic book based on the film. This is exactly what Robert McCall (1919-2010) was called to do before creating the posters, the cover for the soundtrack recording, and the entire iconography of the advertising campaign for Stanley Kubrick's *2001: A Space Odissey*. The existence of this preparatory comic allows us once again to find very evident traces of Bonestell's influence. As proof of this, I will present a comparative analysis of images by Bonestell, Hergé (Hergé, 1953, 1954), and McCall (see Fig. 3). Bonestell's original image shows a chain of mountains surrounding a large crater, which dominates the image. The same mountain chain appears 1) in the Belgian cartoonist's work, 2) in the film's storyboard, and 3) in the record's cover, both created by Mc Call. At the centre of the crater Bonestell drew a round shape, representing the effect of an asteroid's impact. There is a similar shape in the images by both Hergé and Mc Call, marking their visual fulcrum. There is a rocky protuberance on the left-hand side in Bonestell's illustration, and we find the same element in the images by the other two artists. To the right of this protuberance, Bonestell at last introduces a human element. Needless to say, both of the other artists also chose to do so. In other words, if the golden section represented the mathematical model upon which the painters of the Renaissance based the distribution of weights and elements in their paintings, Bonestell's *modus operandi* assumed the same rôle for his fellow creators of space art. This naturally makes the question arise, from a very legitimate doubt: would we have ever reached the moon if NASA's propaganda had been based on Roudaux's more realistic and scientifically correct paintings, rather than on Bonestell's, which were marvellously misleading and so erroneously romantic? One legitimate answer may be: "no, nobody would have invested a penny on an "anemic" mission, which had no visceral appeal on the public". That mission probably would have not taken place until more recent times.

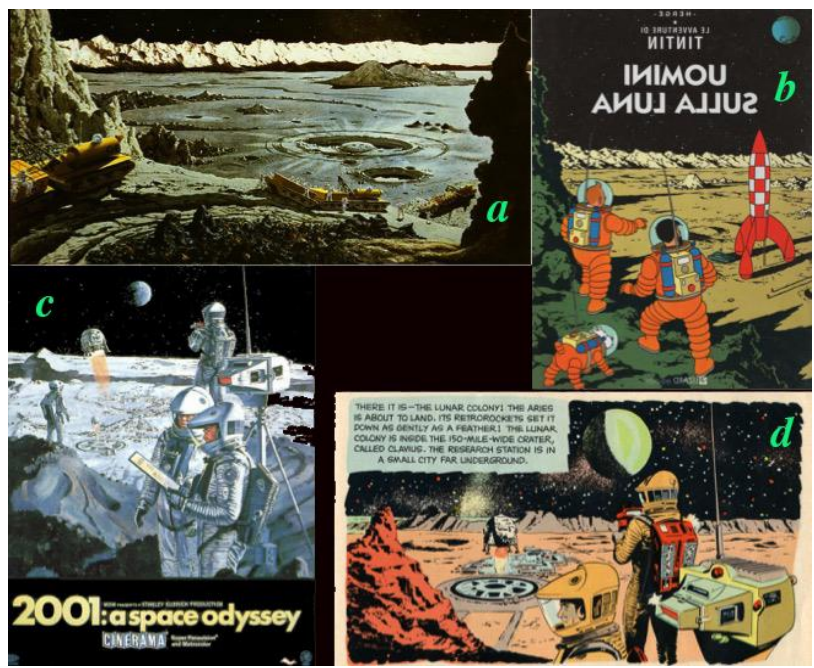

*Figure 3 - Comparison between illustration given by C. Bonestell (a), Hergé (b). As you can see, I have mirrored the latter picture to show in a better way the similarity with Bonestell's illustration and R. Mc Call's (c,d)*

## 4. CHANGING PERSPECTIVES

By analyzing what has been happening under our very eyes since I started mentioning Galilei, Trouvelot, Bolton, Rudaux, Bonestell and Mc Call, I propose a sequence (see Fig. 4) which should make us understand better how our artistic taste has changed through time. In the following, for reasons which will be explained, any mention of artistic taste will be equivalent to discussing a new epistemic requirement. I suggest starting from Galileo Galilei's drawings. In 1609, with his new instrument which will later be called "telescope", he observed various celestial bodies, including the moon. Without the support of photographic instruments, which are now so important and essential for astronomical research, Galileo started a tradition which would last to the end of the 19th century: he made detailed drawings of all the objects he studied, showing very good skills (he had learned from his painter friend Ludovico Cardi, known as *il Cigoli*) and a considerable attention for detail. The well-drawn detail of that historical period will evolve throughout the centuries, and become the modern astronomical datum, accurate and contained in a few *pixels*. Ergo: knowing graphic arts, at the time, meant having one of the essential instruments to practice good science (though the concept of *science* still had to be established). The more accurate the drawing, the higher the scientific content. We should also note that the moon drawn by Galilei is entirely contained in the field of view of his instrument: it can be appreciated in its entirety, and the viewer could distinguish its edges and the details of its surface. While the incomplete drawings of the moon cautiously made a few months earlier by Thomas Harriot (1560 - 1621) – not supported by any writing to clarify the author's thoughts, should any have existed – are considered by some as an expression of maturity of English science, to be seen as a historic preamble to Galilei's observations, I would instead invite to consider the different qualities of the two representations according to my criteria defined earlier: *well executed draw-* *ing, high scientific content – poorly executed drawing, very low scientific content, and no new ideas on reality*. In the end, the correct use of a language is vital for the communication of ideas. If one can not understand a language synthetically, one can not evaluate the information it conveys.

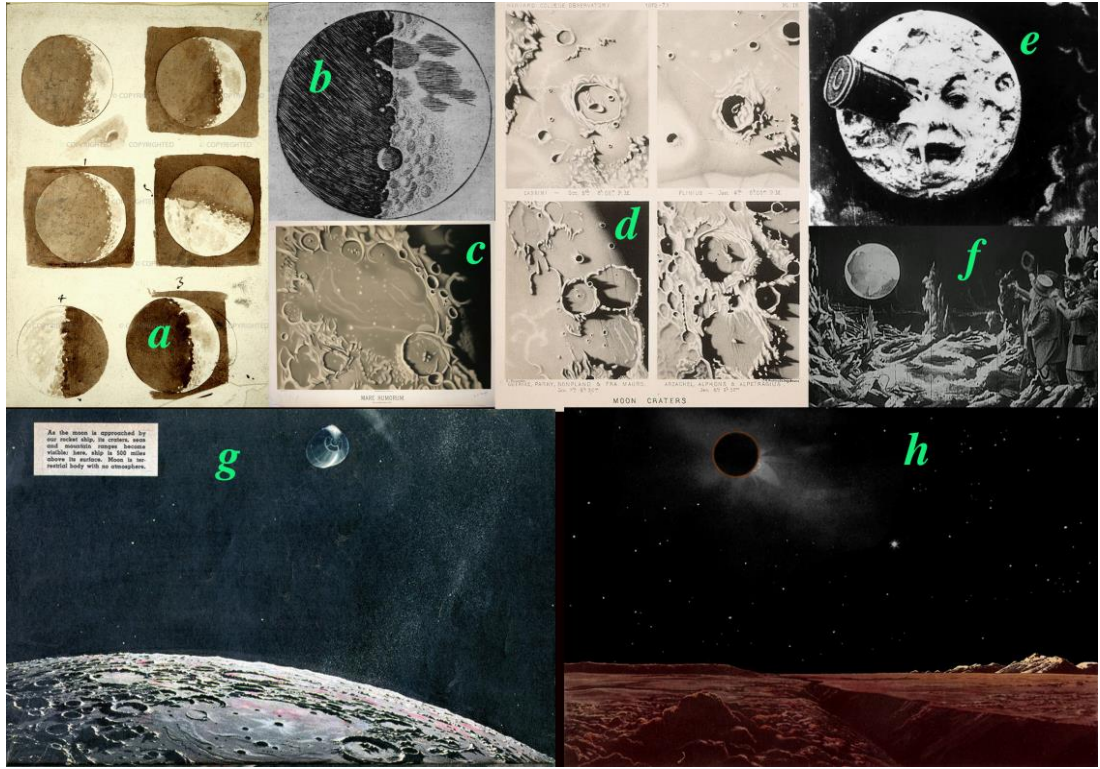

*Figure 4 – Change of the point of view from a realistic one as in a, b by G. Galilei and in c, d by E. Trouvelot to an imagined one as in e, f by G. Méliès (f is to be compared with Bolton's c, d, e illustrations in Fig.1), in g by C. Bonestell (the moon seen from a spacecraft approaching it) and in h by L. Rudaux (an eclipse as seen from the Moon)*

We can evaluate precisely the quantity and quality of Galilei's message without having to make assumptions on the ideas supporting his art, mathematics, and rhetoric. He masterfully used three means of expression: mathematics, writing, and drawing. He made sure to learn the latter from his painter friend for just the same reason that led Einstein to study tensor calculus introduced by Ricci Curbastro: his new reasoning required to use, in order to explain his discoveries, instruments new to him, which he humbly learned.

About two centuries after Galileo, we find the incredible work by the painter and astronomer Etienne Trouvelot, from which we derive, as well as a knowledge of the undisputable artistic skill of the astronomer, also the notion of a much better quality of the instruments: his moon is not seen in its entirety. Only a section of it is masterfully displayed showing some of its details, because the scientist and artist can now use a much higher magnification than that available to Galileo and his primitive instrument.

One could do no better, other than increase the resolution with which the planetary surfaces are observed and mapped. So the next step, taken by Bolton first, and then by Rudaux and Bonestell, does not contemplate the observation of celestial objects from the earth: the following true, great evolution of telescopes will be sending the instruments into space; but the real leap forward, only theorized for now, but realized later, was at the time the fantasy of being there, on those surfaces which so far had been only observed from a great distance, and maybe even glancing towards that "little Earth stage" which we still find so difficult to leave.

Being on the surface of another planet: just fancying this is an *emotional simulation* which, if one uses the techniques described above, is guaranteed to involve the viewer; and while before this time one could imagine to be elsewhere only through literary images (e.g., the works of Ariosto or Verne, to quote just a few), from the early 1900s painted images became available to help the public make this mental leap, which was not an easy one at that time: "I am not on the earth. I am there, on the moon, and from there I am watching the earth". Such a change of outlook was not trivial for that epoch, and, contrary to what we have been led to think, I do not believe that it started with Bolton.

## 5. ORIGIN OF THE IDEA

I think instead that the birth of this tendency to see the cosmos from places other than the earth can be placed at least as far back as 1902, when the famous *Voyage dans la Lune* by Melies (1861-1938) was shown for the first time. It should be noted that, at the time, Bolton was only 19 years old. In that movie, the moon looks around, receives a manned bullet in a eye and cries for the pain. It was very simple for the public to engage with her while she looks to the universe from her point of view and when she cries. The surface of the moon, represented as a vulcanic world, is completely the same as that painted by Bolton (see Fig. 1 and Fig. 4).

The character of these operations is, I believe, much more complex, and claims some heuristic characteristics which, without available instruments, become or have reasonable expectancy to become scientific characteristics.

## 6. ANALOG SIMULATION

If we are prepared to receive these illustrations as *emotional simulations*, perhaps we should try to go further and accept them as *analog simulations*, or quite simply as simulations: no matter how far we will be able to reach in the future with human spaceflight, visiting new worlds, there will still be places beyond the reach of our technology. The only available way to imagine how they look like, and how the universe appears from there, will be through an artistic representation: but it will have to be *educated*, guided by a strong scientific support which prevents it from going where artistic vision alone could dangerously lead it. The "physical leash" to impose on the paintbrush is a guarantee, albeit a tenuous one, of moving within acceptable error bars. And we know from cosmology that "acceptable" could mean as much as an error of 20% on the measured quantity.

After all, today's computer simulations of environments which can not be experienced directly are based on a physical picture which, we know, works well "here": we extend it to "there" on the basis of the so-called "principle of uniformity of nature". This is similar to recreating distant planetary environments by painting an earth landscape, leaving out the flora and adding stars and craters to the remaining geology. I will not discuss a comparison between the emotions evoked by a painting and those induced by contemplating a computer representation of distant worlds; but I invite the reader to consider the possibility of "vibrating" esthetically, and possibly reaching great emotional involvement, even when facing a Montecarlo simulation. This I consider certain: just like, before the invention of adequate cameras, the only means of recording observations were drawing and painting, similarly, before the advent of powerful computers and programs, the only way to achieve physical simulations was through artistic representations based on an altered point of view.

A point of view which, when we observe the mixture of stars, gas and planets which we call Galaxy, is blatantly *Lagrangian*; but when we ideally place ourselves elsewhere to remeasure what we see in the sky from a new location, it shows a *Eulerian* tendency to photograph the field, or rather, to draw it.

## 7. THE SOCIOLOGICAL ASPECT

At this point, I would like to discuss something which I consider very important. So far, we have only considered the impact of such illustrations of the cosmos on society, without discussing how they were considered within that very scientific community of which they were officially an expression. The historians Miller and Durant III (Miller and Durant III, 2001) recall how, once the images from lunar surveyors had revealed that the lunar surface had been modelled and smoothed for billions of years by meteor impacts, and was much less interesting than its interpretation by Bonestell, the scientific community at first seemed reluctant to accept the facts, and instead seemed more inclined to believe that our satellite looked as the artist had imagined it. Naturally, all the photographic evidence quickly caused a change in the trend, but at first even scientists were prone to believe what they were more used to, a habit created by the deep emotions which those images evoked. This might be an uncomfortable truth, but in those circumstances many of NASA's scientists, far from impersonating the cool and detached cliché, turned out to be… people. Ultimately, we should remember how widely Bonestell was accepted in a scientific community which was (and to some extent still is) rather closed, and was just timidly starting to open up, but only for political reasons. And its opening came more from duty than conviction, to an outside world which had become important, and had to be convinced, because of the financial support the scientific world needed for its largest and most ambitious projects. The sociologist of science Massimiano Bucchi (Bucchi, 2003) reports a relevant fact in his essay:

*A further event of great significance took place in 1957. In that year, the Soviet Union launched the first artificial*

*satellite in history; a feat which had enormous impact in the Western countries, and especially in the US, where the launch was considered indicative of the progress – and therefore the dangerous potential – achieved in science and technology by the rival superpower. The 'Sputnik effect' triggered reactions at two levels. First, there was a further expansion in research expenditure in the US, which grew by around 15 per cent per year until the early 1960s. At the same time, politicians and scientists became convinced that competition with the Soviet Union could only be sustained if greater commitment was made to university education and, in particular, to the training and recruitment of advanced researchers and technicians.*

Therefore, I believe that Bonestell's protection by the scientific environment was connected to his media exposure and that of his work, which was obtained through his previous experience as a painter of *mattes* for films, and later through his achieving a strong position as a NASA collaborator. I am essentially describing an *ante litteram impact factor*, which can be translated as a sort of *authority principle* which favoured the American illustrator after all his work on space subjects for the large research institutions, and even side-to-side with the father of the *Apollo* project, Werner von Braun. This throws a rather oblique light on the scientific community of that time which, rather than feeling bound by the fundamental principles so often stated by modern epistemology, often appeared dynamically sensitive to the overly human trends displayed by *those who work as scientists*. The "normative structure of science" (Merton, 1942) which could guarantee an ideal functioning of the scientific community if adopted, was based on four famous, fundamental imperatives: universalism, "communism", disinterestedness, and organized skepticism. We again turn to Bucchi:

*In the early 1970s, several studies, including a detailed survey of forty-two scientists involved in analysis of data on the moon's surface collected by the Apollo missions, sought to demonstrate that such ambivalence gave rise to a 'dynamic alternation of norms and counter-norms' (Merton and Barber, 1963: 104). The institutional imperatives enunciated by Merton were thus matched by counter-norms such as 'particularism, interestedness, and organized dogmatism' (Mitroff, 1974). The scientists interviewed by Mitroff, for instance, attributed the following characteristics to themselves and to their colleagues: a reluctance to make certain aspects of their research public; an attachment to their own hypotheses; an unwillingness to abandon these hypotheses even in the presence of data contrary to them; or the tendency to judge results and claims on the basis of the social attributes (nationality, academic position) of the scientist advancing them.*

A point-by-point analysis of table 1 reveals how the scientific community, in Bonestell's case, routed around its artist, defending him just like any other corporation would do with one of its illustrious members, contradicting every single point of Merton's programme of research.

## 8. CONCLUSIONS

As I have noted elsewhere (Adamo, 2009), ultimately art and science provide representations of our reality which can, and should, cooperate. I am also sure that this cooperation already exists, but is still minimal, especially because of science's reluctance to accept that these interactions are inevitable. Unfortunately, as acting parties in the process, we cannot be fully aware of what we are doing or not doing, but this will not save us from the judgement of those who will study our actions in the future. I would end with some provocative questions. After asking ourselves what the result would have been if Rudeaux's more objective but less interesting paintings had been chosen to promote space exploration, I ask myself and you:

1) knowing what science is, what goals it should achieve, and how it should achieve them, are we sure that scientists should always be "honest" in their interaction with a society formed by people who support research through taxes? Should they really avoid feeding the public with beautiful illusions, preferring instead to distribute scientific exactness, which is fascinating only for insiders?

2) Are we sure that scientists should always communicate with the public? Would it not be better to let this be done by trained communicators, supported by a solid scientific background?

And finally:

3) Are we sure that science is that different from art?

*Table I. Merton's Norm VS Counter-Norms from Mitroff's study*

| Norms | Counter-norms |
|---|---|
| **Universalism** Scientific claims and findings are judged independently of the person and social attributes of their proponents: social class, race, religion. | **Particularism** A Scientist's social characteristics are factors which importantly influence how his/her work will be judged. |
| **Communism** Findings and discoveries are not the property of the individual researcher but belong to the scientific community and to society at large. | **Individualism** Property rights are extended to include protective control over results. |
| **Disinterestedness** Scientists pursue their primary aim, knowledge progress and indirectly achieve individual rewards. | **Interestedness** The individual researcher seeks to serve his/her own interests and those of the restricted group of scientists to which he/she belongs. |
| **Organized scepticism** Every researcher is obliged to scrutinize every hypothesis or finding carefully, including his own, suspending final judgement until the necessary confirmations become available. | **Organized dogmatism** The scientist must believe in his/her own findings with utter conviction while doubting those of others. |

## ACKNOWLEDGEMENTS

I greatly thank my colleague Giovanna Stirpe, astronomer at INAF-Astronomical Observatory in Bologna (IT), who has kindly translated this long article from Italian.